\documentclass[journal]{IEEEtran}

\usepackage{ifpdf}
\ifCLASSINFOpdf
  \usepackage[pdftex]{graphicx}
  \DeclareGraphicsExtensions{.pdf,.jpeg,.png}
\else
  \usepackage[dvips]{graphicx}
  \DeclareGraphicsExtensions{.eps}
\fi

\usepackage[utf8]{inputenc}
\usepackage[T1]{fontenc}
\usepackage[french, english]{babel}
\usepackage{cite}
\usepackage{caption}
\usepackage{amsmath,amsfonts,amssymb}
\usepackage{subcaption}
\usepackage{array}
\usepackage{color}
\usepackage{float}
\usepackage{multicol}
\usepackage[]{algorithm2e}
\usepackage{hyperref}
\usepackage{fancyhdr}
\graphicspath{{Figures/}}

\begin{document}
\pagestyle{plain}
\pagenumbering{arabic}

%
%
\title{Stability and Performance of Coalitions of Prosumers Through Diversification in the Smart Grid}

%
%
\author{Nicolas Gensollen, Vincent Gauthier, Monique Becker and Michel Marot  \\
\IEEEauthorblockA{CNRS UMR 5157 SAMOVAR, \\
Telecom SudParis/Institut Mines Telecom\\
Email: \{nicolas.gensollen, vincent.gauthier, michel.marot, monique.becker\}@telecom-sudparis.eu}}

\maketitle

%
%
\begin{abstract}
Achieving a successful energetic transition through a smarter and greener electricity grid is a major goal for the 21st century. It is assumed that such smart grids will be characterized by bidirectional electricity flows coupled with the use of small renewable generators and a proper efficient information system. All these bricks might enable end users to take part in the grid stability by injecting power, or by shaping their consumption against financial compensation. In this paper, we propose an algorithm that forms coalitions of agents, called prosumers, that both produce and consume. It is designed to be used by aggregators that aim at selling the aggregated surplus of production of the prosumers they control. We rely on real weather data sampled across stations of a given territory in order to simulate realistic production and consumption patterns for each prosumer. This approach enables us to capture geographical correlations among the agents while preserving the diversity due to different behaviors. As aggregators are bound to the grid operator by a contract, they seek to maximize their offer while minimizing their risk. The proposed graph based algorithm takes the underlying correlation structure of the agents into account and outputs coalitions with both high productivity and low variability. We show then that the resulting diversified coalitions are able to generate higher benefits on a constrained energy market, and are more resilient to random failures of the agents.
\end{abstract}

\IEEEpeerreviewmaketitle

%
%
\section{Introduction}
\label{sec:introduction}

Designing stable power systems is a classical engineering challenge since blackouts can have catastrophic consequences. One obvious condition for stability revolves around sustaining at any time a power generation that meets the demand. If one is larger than the other, the system deviates from its synchronous state. If nothing is done to return to the synchronized equilibrium, this can lead to catastrophic cascades of failures \cite{Brummitt2012} \cite{Wang2009}. With traditional power plants based on fossil energies, the production can be scheduled in order to sustain the predicted consumption. Deviations from this schedule can then be supported almost on real time by using fast response power plants, interconnections with border countries, as well as regulator entities. Even if most individual consumers may not realize it, the electricity prices are not constant rates and evolve with the production/consumption conditions on the grid. It is often assumed that the granularity of these prices is meant to increase in the future as a way to pass on the production conditions on the end users \cite{Jiang2014}. The pricing of electricity and the necessity for the grid operator to have different emergency reserves lead to an economy setting for electricity, where operations and various kind of contracts are decided on a market \cite{Europe}. Such a market environment clearly necessitates communication in order to monitor and obtain information from the grid as well as exchange between participants. These kind of settings are already being used at the transport level of current power grid with large renewable plants in the production portfolio. Using electricity markets appears thus as a way to manage the reliability of the whole system, especially when the use of renewables is important.

The smart grid vision however goes further and revolutionizes this top-down centralized architecture by assuming that bidirectional power flows are allowed. This would change completely the nature of the traditional grid since production could be located down to the very end of the distribution networks. Nodes that were simply pure loads yesterday could behave tomorrow as generators or loads \cite{Ramchurn}. On the other hand, the user of renewables in the production is continuously growing, and is expected to become majority in a near future. These plants completely rely on the presence and the intensity of their respective resources (wind, sun, tiles...). Balancing production and demand in such a scenario appears much more challenging and it seems inevitable that the grid should modernize its infrastructures to sustain this transition \cite{Europe}. The key to drive safely such a system is assumed to be information, and more precisely, the capacity to measure, communicate, and analyze data on real time.

In this paper, we focus on these "\textit{nodes}" in the distribution network that can produce and consume electricity. More precisely, we consider a set of agents that own both renewable distributed energy resources (DER) and electrical loads. The production of an agent can be, of course, used to meet its own demand, but in cases when it is over-producing, we consider that it has the possibility to inject, against financial compensation, its extra-production in the grid. Such an agent model is known as a "\textit{prosumer}" \cite{Rathnayaka2012} (and will be called accordingly in this paper). A key point of this work is to merge the interests of the grid operator with the prosumers in a single utility function. While the latter intend to maximize their benefits with higher production contracts, the operator primary concerns are related to the quality and stability of the injections. The coalitions that we wish to form should thus be both stable and productive. This compromise can be difficult to obtain in real situations, especially when renewable generation and complex spatiotemporal correlation happen.

More precisely, in our model, any entity that wishes to sell its production on some energy market has to estimate and announce a production capacity for an upcoming period of time. If a contract is bought, the entity commits to injecting exactly, at any time of the contracted period, the announced amount of power, under financial penalties if it fails. Since prosumers use exclusively renewables, contracts come naturally with some amount of risk due to the intermittent nature of most renewables. Using storage as buffers in order to reduce risks is a popular approach. Actually, a whole branch of the smart grid literature is even studying the possibility to use electric vehicles as moving storage capacities for stabilizing the grid. Without storage and proper control, the over-producing state of the agents, and thus the power they inject in the grid, might be rather unstable. This is clearly unacceptable for the grid operator that cannot ensure system stability if it has to deal with numerous small unpredictable entities.

Forming coalitions/aggregations is an envisaged solution to both the number of entities soliciting the operator, and their high variability. Indeed, it has been argued that using a multi-level aggregators architecture for the control could lead to better performance on the communication side \cite{Negeri2012}, since an aggregator can be considered as a single node by the level above it. On the other hand, it is well-known that diversification of the assets is a way of minimizing risks when constructing a portfolio. One thus expect a more stable and predictable energy production for a coalition than for a single agent. Nevertheless, all coalitions are obviously not equally stable or productive, which means that special attention should be paid to the aggregation step. Recall that prosumers have both a consumption and a production component, each depending on location and time, meaning that there are complex underlying correlations between the agents. This is a central topic of the present paper : given N prosumers, what coalitions should be formed so that the compromise between expected production and variability is optimized ?

We will see that the variability in the aggregated productions can be quantified to a certain extent by the correlation among the agents forming the coalitions. Understanding the correlation relationships among the agents can thus give an indication about what coalitions to form and how much they should sell. More precisely, we use a graph representation of the correlation structure to gain insight about the expected production to risk ratios of different coalitions. We build a framework in which the system operator specifies both the minimum production acceptable to enter the market and restrain the amount of risk he is willing to take. We then propose a graph heuristic that uses decorrelated cliques in order to form diversified productive and stable coalitions, and we compare the results with other formation strategies (see section \ref{sec:results}). 

Rather than maximizing the profit of each agent in the grid (prosumers and grid operator), our algorithm tries to form the most productive coalitions given a maximum amount of risk acceptable. In other words, it tries to maximize coalitions profits without considering individual retributions to single agents (a task often studied through game theory). Furthermore, the pricing strategies for both the prosumer and the grid is beyond the scope of this paper.  

Because agents are susceptible to fail for diverse reasons, the propensity of the system to undertake these failures is critical \cite{Pahwa}. We therefore investigate in this paper the resilience of the coalitions when prosumers fail. Despite the fact that losing agents is usually detrimental to the coalitions, we will see that the coalitions formed with our algorithm tend to be less impacted by random failures.

The paper is organized as follows, section \ref{sec:related} gives a brief overview of the related literature, section \ref{sec:data} clarifies how we generated realistic prosumer production traces based on weather data. In section \ref{sec:notations}, we define most of the notations and explain why correlation between prosumers is a quantity of interest for our objective. Based on the conclusions of section \ref{sec:notations}, we develop in section \ref{sec:utility} a utility function quantifying how much power a coalition can announce on the market given an accepted risk level. We then develop, in section \ref{sec:forming} a greedy optimization algorithm that uses decorrelated cliques as inputs and improves their utility over a correlation-constrained environment. Finally, section \ref{sec:results} provides some results both on performance of the method and resilience of the coalitions formed.

%
%

\section{Related Work}
\label{sec:related}

As explained in the section above, allowing entities using mainly renewables to inject power in the grid is a difficult challenge. Indeed, on the contrary to fossil plants whose production can be scheduled in advance to meet the expected consumption, renewables by definition only produce when the resource is present. Unfortunately, these moments do not necessarily coincide with the consumption peak hours \cite{Milligan2010}. One possible solution consists in shifting some of the loads towards high productivity periods. Demand side management techniques implemented in the end users' smart meters, can mitigate the consumption peaks and wast less production \cite{Milligan2010} \cite{Logenthiran2012}. Since end users tend to seek a maximum utility at a minimum cost, dynamic pricing is believed to be a good way to give incentives to them. By carefully scheduling the prices, it is assumed that the load curve can be shaped to some extend.

Dynamic pricing is an interesting and useful tool, but it has also some limitations. It is likely that dynamic pricing will serve as a shaping mass tool while finer techniques will be needed locally. A popular approach in this direction consists in deploying storage devices in the network and using them as electricity buffers. Basically these storages would be charged when there is a surplus of production, and discharged when the consumption exceeds the production. Although quite simple, this idea causes numerous challenges when it comes to its real implementation. As long as the system considered remains small, a centralized control of these buffers could be envisioned. However, for real large systems it is likely that more sophisticated decentralized algorithms will be necessary. In \cite{Zhang2013}, the authors introduce a distributed energy management system with a high use of renewables such that power is scheduled in a distributed fashion. In \cite{Shadmand2014} the optimal storage capacity problem is addressed. There is indeed an interesting trade-off between the costs of the equipments and the expected availability of power. The authors develop a framework that enables them to exhibit a Pareto front of efficient solutions.

These are only a few possibilities for balancing production and demand in the smart grid. Most of the time, these techniques will be coupled with predictions of the upcoming load curves and weather conditions. Combining all these technologies enables aggregators to quantify their expected production and the inherent risk that comes with it. The optimization of expected returns to risk is a traditional goal in finance, and a wide literature exists on this topic. It is well-known for instance, that the more risk one is willing to take, the higher his potential gains. On the contrary, when investing exclusively on low risks assets, one should expect relatively small gains. This trade-off is formalized in the Markowitz' portfolio theory \cite{Markowitz1959}. More precisely, given a set of assets for which we have some historic data of returns, the objective is to find a linear combination of these assets (the so-called portfolio) which maximizes the expected value while minimizing the variance of the portfolio's return. Markowitz's answer is a set of efficient portfolios that all optimize in some sense this trade-off. If one is able to put a number on his risk acceptance or on the target expected return, the corresponding efficient portfolio is a priori the best option. One of the most controversial assumptions in the portfolio theory is that returns are jointly normally distributed (or, at least, that the returns distribution is jointly elliptical). Some economist have pointed out the fact that this assumption might not capture well the reality of financial markets \cite{doi:10.1142/S0219024912500197}.

Nevertheless, one of the key point in the Markowitz theory is to consider explicitly the correlation between the assets since they impact directly the variances of the portfolios. Since the work of \cite{Mantegna1999}, an interesting approach consists in computing a distance metric based on the correlation coefficients in order to organize the series in a correlation graph. Nodes represent the series considered while the edges are weighted by the metric. Because the metric can be computed for all pairs, these graphs are complete and of little use as is. Historically, the approach used by \cite{Mantegna1999} was to compute a minimum spanning tree as to obtain a hierarchical clustering of the series. Later on, it was pointed out that, by definition, a spanning tree could not capture the underlying clustering structure hidden in the correlation graph. In this paper, we use another classical filtering technique called $ \epsilon $-graph \cite{Garas2008}. It consists in selecting a threshold $ \epsilon $, and filtering out edges with smaller weights. As we will see further in this paper, this approach has the advantage of preserving clusters of correlated series.

%
%

\section{Generating realistic prosumer patterns}
\label{sec:data}

An essential component of the smart grid is the smart meter which makes the interface between the end user and the rest of the system. Smart meters coupled with sensors measure quantities of interest (like instantaneous consumption), receive informations from the grid (electricity prices for instance), and take actions accordingly (demand side management programs). Smart meters are currently and gradually deployed, and will probably provide interesting datasets to work on. Unfortunately, at the time this paper was written, production and consumption data for prosumers over a large region were not yet available to our knowledge. Some interesting experiments are nonetheless being conducted and data are progressively made public \cite{ISSDA}. 

In this paper, we use weather quantities like wind speed or solar radiance as alternative data for generating realistic production and consumption series. Fortunately, these kinds of data are easier to find, and since the development of small personal weather stations, their geographical granularity keeps increasing.  Since these quantities depend both on time and location, we discretize time into slots and space into zones in the following. A zone is simply a portion of the considered region of study for which we sampled data. Therefore, if prosumers i and j are positioned on the same zone, they are exposed to the same weather. Adding some intra-zone noise can easily be done though not considered in this paper.

More formally, we denote by $ P_{i}(t) $ the instantaneous extra-production of agent i at time t :

\begin{equation}
P_{i}(t) = P_{i}^{P}(t) - P_{i}^{D}(t)
\end{equation}

Where $ P_{i}^{P}(t) $ represents the total production of agent i at time t and $ P_{i}^{D}(t) $ its consumption at time t. In other words, $ P_{i}(t) $ represents the instantaneous surplus of power that agent i is willing to sell at time t. As explained above, since large datasets containing this quantity over time are not yet available, we simulated these traces by considering separately $ P_{i}^{P} $ and $ P_{i}^{D} $.

For a prosumer i, it is possible to write both quantities as a sum over the distributed energy resources ($ DER_{i} $) and loads ($ load_{i} $) of i : 

\begin{equation}
P_{i}^{P}(t) = \sum_{k \in DER_{i}} P_{k}(t)
\end{equation}
\begin{equation}
P_{i}^{D}(t) = \sum_{k \in load_{i}} P_{k}(t)
\end{equation}

For simplicity, in this paper we only consider wind-turbines (WT) and photovoltaic panels (PV) as possible DERs for the agents ($ DER_{i} = WT_{i} \cup PV_{i} $):  

\begin{equation}
P_{i}^{D}(t) = \sum_{k \in WT_{i}} P_{k}(t) + \sum_{k \in PV_{i}} P_{k}(t)
\end{equation} 

\begin{figure}
\includegraphics[scale=.49]{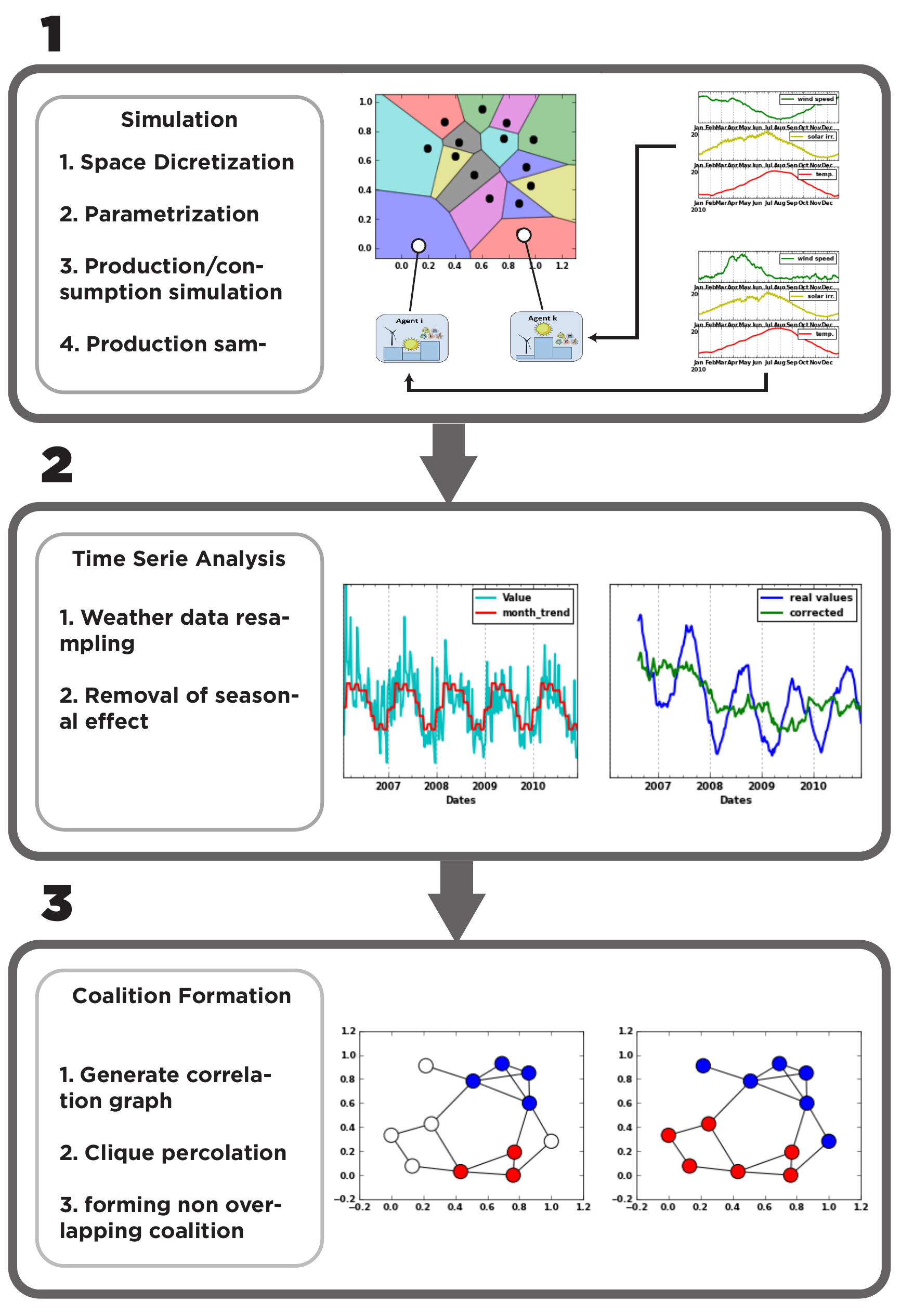}
\caption{Process diagram}
\label{fig:process}
\end{figure}

We denote by $ \nu_{i}(t) $ and $ \Psi_{i}(t) $ the wind speed (in $ m.s^{-1} $) and the solar radiance ( in $ W.m^{-2} $) at the location of agent i and at time t, so that :
\begin{equation}
 P_{i}^{P}(t) = \sum_{k \in WT_{i}} \mathcal{F}_{WT}( \nu_{i}(t) ) + \sum_{k \in PV_{i} } \mathcal{F}_{PV}(\Psi_{i}(t) ) 
\end{equation}
Where $ \mathcal{F}_{WT} $ (resp. $ \mathcal{F}_{PV} $) is the power curve for the wind-turbines (resp. photovoltaic panels). We made here the implicit assumption that all wind-turbines (resp. photovoltaic panels) have the same power curve. The model can be easily extended to multiple power curves accounting for different types of generators. More details about power curves and their approximations can be found in the appendix \ref{appendixB} and in \cite{Lydia2014}. The process for generating the $ P_{i} $ series is pictured in the first block of the process diagram (see figure \ref{fig:process}).

Note that a prosumer i is defined by his zone $ Z_{i} $ as well as the sets $ DER_{i} $ and $ load_{i} $. That is, a prosumer can be configured to represent anything from a single wind-turbine for instance ($ DER_{i} = \{ WT_{0} \} $ and $ load_{i} = \emptyset $) to a pure load ($ DER_{i} = \emptyset $ and $ load_{i} = \{ L_{0} \} $) through more complex combinations. In practice, we use random configurations for the agents.

In the rest of the paper, we use french weather data \cite{Infoclimat} starting in January 2006 and ending in December 2012, with a sampling frequency of three hours, and generate N timeseries of extra-production over this date range.

%
%

\section{Notations}
\label{sec:notations}

This section provides most of the notations and introduces important concepts for the rest of the paper. As explained in section \ref{sec:data}, we consider a set $ \mathcal{A} = \{a_{1},a_{2},...,a_{N} \} $ of N prosumers configured randomly, and for each agent, we simulate its extra-production $ P_{i}(t),\ \forall i \in \mathcal{A} $ from 2006 to 2012. Based on these historical values, our objective is now to form groups of prosumers (the so-called coalitions) so that the global power production resulting from the superposition of individual's extra-productions be both sufficiently high and predictable. Let $ P_{S}(t) = \sum_{i \in S} P_{i}(t) $ be the extra-production of coalition S at time t. 

Suppose now that coalition S has to suggest a production value $ P_{S}^{CRCT} $ to enter the market. This means that, during the time S is on the market, it will have to inject in the grid exactly $ P_{S}^{CRCT} $ at any time t and will be rewarded proportionally to this amount, with penalties if it deviates. Obviously, the actual extra-production will not be constant at this value and will oscillate due to intermittences in the production and consumption. If S always produces more than $ P_{S}^{CRCT} $, it will never have to pay penalties, but it is losing some gains since it could have announced a higher contract value. If the production oscillates around $ P_{S}^{CRCT} $, by using batteries or demand side management techniques (see section \ref{sec:related}), S could be able to maintain its production to the contract value at any time. Nevertheless, if the oscillations are too important compared to the available storage capacity, S will probably break the contract and pay penalties. We can see that there is a return over risk trade-off here, meaning that coalitions should find the right balance between announcing too low and losing some potential gains, and claiming too high and paying penalties. 

Let us illustrate the rest of the notations and concepts with a simple example. We consider only two agents i and j such that the distribution of their extra-production can be approximated by normal distributions : $ P_{i} \sim \mathcal{N}(\mu_{i}, \sigma_{i} ) $ and $ P_{j} \sim \mathcal{N}(\mu_{j}, \sigma_{j} ) $. This is only for explanation purposes as it is of course rather unrealistic in real situations where the distributions are skewed. Using simple statistics, we can write the distribution of the coalition $ S = \{i,j\} $ as $ P_{\{i,j\}} \sim \mathcal{N}(\mu_{ij}, \sigma_{ij}) $, where :

\begin{equation}
\left\{ \begin{array}{lll}
		\mu_{ij} = \mu_{i} + \mu_{j} \\
		\sigma_{ij} = \sqrt{\sigma_{i}^{2} + \sigma_{j}^{2} + \rho_{ij} \sigma_{i} \sigma_{j} }
\end{array} \right.
\end{equation}
$ \rho_{ij} $ being the Pearson's correlation coefficient between $ P_{i} $ and $ P_{j} $. If the coalition $ \{i,j\}$ proposes a contract value $ P_{S}^{CRCT} $, all instants when $ \{i,j\}$ will produce less than $ P_{S}^{CRCT} $ is critical. Indeed, in this kind of situations, $ \{i,j\}$ will either have to discharge batteries to keep up with its contract, or pay penalties to the grid. The probability that $ \{i,j\}$ is under-producing compared to the contract : $ Pr[P_{i,j} \leq P^{CRCT}] $ is thus an important indicator of the coalition's quality. A well-known result for normal distributions is that the cumulative distribution function can be written as :
\begin{equation}
Pr[P_{ij} \leq P_{S}^{CRCT}] = \dfrac{1}{2} \left[ 1+ erf \left( \dfrac{P_{S}^{CRCT} - \mu_{ij}}{\sigma_{ij}\sqrt{2}} \right) \right] 
\end{equation}
where $ erf $ is the error function : $ erf(x) = \dfrac{2}{\sqrt{\pi}}\int_{0}^{x} e^{-t^{2}} dt $.

The contract a given coalition is willing to take depends on its capacity to compensate for under-producing (using batteries, backup generators...), and its risk acceptance. Selecting the right contract value appears thus as an interesting problem on its own that we plan to investigate in future works. In order to keep the present paper in a reasonable length, we simplify the contract value selection problem by giving some responsibilities to a third party named the grid operator. The role of the grid operator is to constrain the market entry to coalitions able to propose both sufficiently high and sufficiently credible contract values. More formally, let $ \phi \in [0,1] $ be the reliability threshold fixed by the grid operator as a maximum value for the probability of under-producing. The highest contract value that a coalition can propose is thus $ P_{S}^{CRCT \star} $ such that $ Pr[P_{ij} \leq P_{S}^{CRCT \star}] = \phi $. In the Gaussian example, it implies that coalition $ \{i,j\}$ is announcing :

\begin{equation}
\label{eq:P_star}
P_{S}^{CRCT \star} = \mu_{ij} - \sqrt{2} \sigma_{ij} erf^{-1}( 1 - 2 \phi )
\end{equation}

\begin{figure}
\includegraphics[scale=.6]{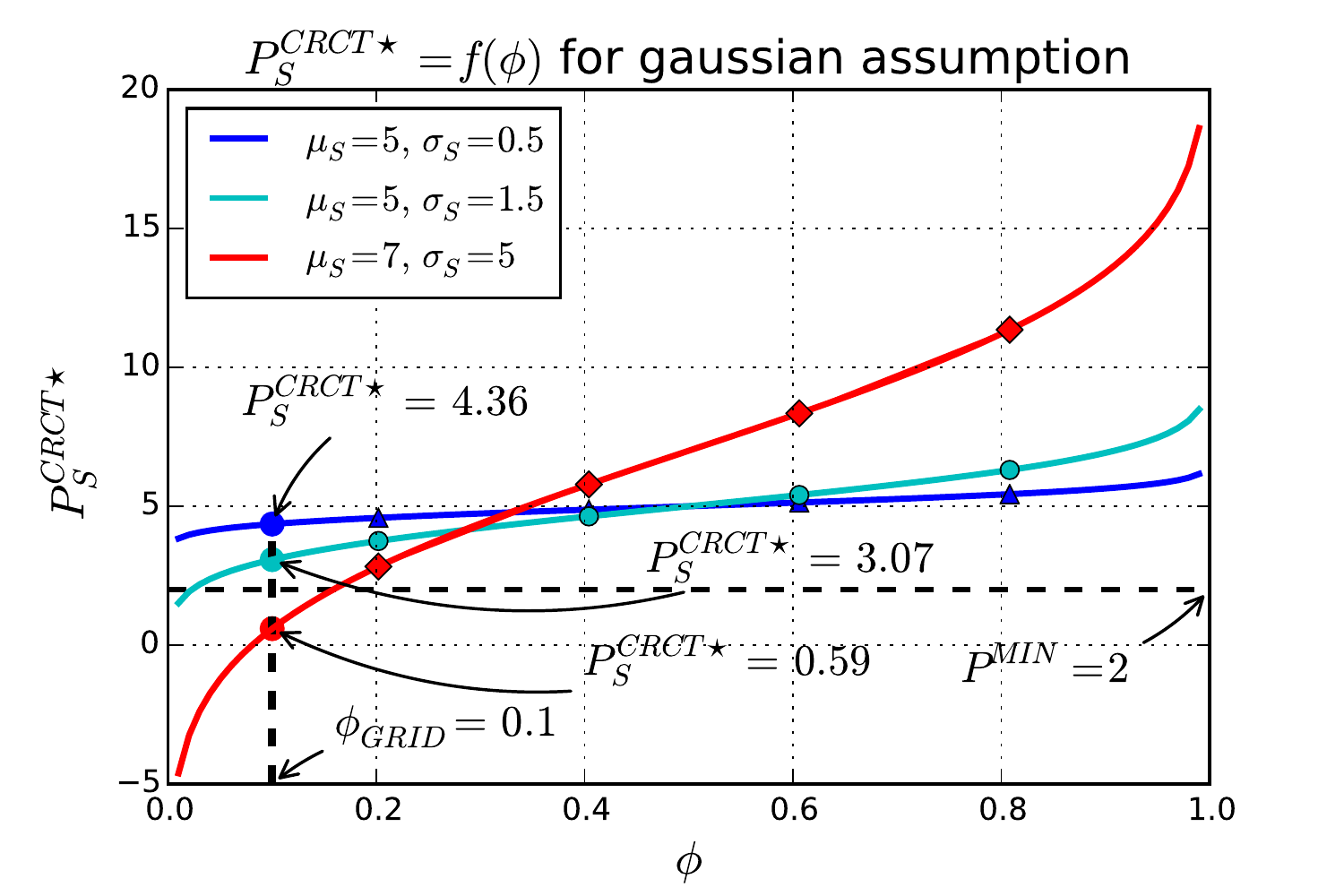}
\caption{$ P_{S}^{CRCT \star} $ depending on reliability parameter $ \phi $ for Gaussian distributions (see equation \ref{eq:P_star}). Blue curve with triangles stands for a coalition S with an expected production of 5 units and a standard deviation of 0.5. Under a grid policy of $ \phi = 0.1 $, it is able to announce a contract value of $ P_{S}^{CRCT \star} = 4.36 $. The same coalition in term of expected production ($\mu = 5 $), but with a higher variance ($ \sigma = 1.5 $, cyan curve with circles) can only afford a smaller contract value of $ P_{S}^{CRCT \star} = 3.07 $. The red curve with diamonds stands for a coalition with a higher expected production $(\mu = 7)$, but with a very high unpredictability ($ \sigma = 5 $). For low values of $ \phi $, this coalition is thus heavily penalized and can only afford a contract of $ 0.59 $ units. Under grid policy $ (\phi = 0.1, P^{MIN} = 2) $, this last coalition is thus not allowed to enter the market (red dot below the horizontal dashed line). }
\label{fig:Gaussian}
\end{figure}

This is the best contract value that the coalition S can afford giving the stability policy $\phi$ of the grid operator. Figure \ref{fig:Gaussian} shows how $ P_{S}^{CRCT \star} $ evolves according to the reliability parameter $ \phi $. For illustration, the range of $ \phi $ values is shown from 0 to 1, but in practice, only small values of $ \phi $ really make sense : $ \phi = 1 $ for instance means that coalitions can announce absolutely anything since the probability of producing less than any contract value is necessarily less than one by trivial definition of a probability. As visible on figure \ref{fig:Gaussian}, coalitions with high expected productions but presenting a high unpredictability are penalized and can only afford small contracts.  

In order not to overload the market with unrealistically small coalitions, the grid operator also specifies a lower bound $ P^{MIN} $ on the contract values. We thus characterized a valid coalition as one satisfying the two conditions :

\begin{equation}
\left\{ \begin{array}{lll}
			Pr[P_{ij} \leq P_{S}^{CRCT}] \leq \phi \\
			P_{S}^{CRCT} \geq P^{MIN}
\end{array} \right.
\end{equation}

On figure \ref{fig:Gaussian}, $ P^{MIN} $ is fixed to 2 units for illustration purpose. For $ \phi = 0.1 $, only blue triangles and cyan circles coalitions are valid while red diamonds coalition is not.

The Gaussian assumption of this small example is convenient as it allows us to write $ P_{S}^{CRCT \star} $ analytically. Nevertheless, such assumption is rather unrealistic in practice. In the following, we keep the same framework but release this Gaussian assumption unless the contrary is specified (see eq. \ref{eq:alpha_star}). This assumption will indeed be convenient for computing some parameter estimates.

%
%

\section{Utility Function}
\label{sec:utility}

In this section, we use the notions of contract values and valid coalitions developed in section \ref{sec:notations} in order to design a proper utility function. The contract basically indicates the rate at which a coalition has to inject power in the grid. It seems then natural that coalitions are remunerated proportionally to their contract values $ gain(S) \propto P_{S}^{CRCT \star} $. More precisely, if $ \lambda $ is the unitary price rate for electricity, a coalition S injecting $ P_{S}^{CRCT \star} $ in the grid during a period $ [t_{0},t_{k}] $ earns :

\begin{equation}
gain(S) = \int_{t_{0}}^{t_{k}} \lambda P_{S}^{CRCT \star} dt = P_{S}^{CRCT \star} \int_{t_{0}}^{t_{k}} \lambda dt
\end{equation} 
(since $ P_{S}^{CRCT \star} $ is supposed to be a constant rate over the contracted period). Using $ gain(S) $ directly as a utility function suffers a major drawback. It is indeed not a concave function of the coalitions' sizes, meaning that coalitions can grow as large as the number of agents allows it, without any counterbalance effects. 

Such a model, that virtually allows infinitely large coalitions and contract values, is in practice not realistic. There are indeed costs (communication costs for instance) that increase with the coalitions sizes. We take this observation into account by rescaling the utility of a coalition S by its size in term of number of agents ($|S|$):

\begin{equation}
\mathcal{U}(S) = \left\{ \begin{array}{lll}
							\dfrac{1}{|S|^{\alpha}} \dfrac{ P_{S}^{CRCT \star} }{P^{MAX}},\ if\ S\ is\ valid, \\
							0,\ if\ S\ is\ not\ valid
						 \end{array}
				  \right.
\end{equation}
where parameter $ \alpha $ controls to what extent the size of a coalition impacts its utility, and $ P^{MAX} $ is a normalizing factor. $ P^{MAX} $ can be seen as the maximum production which can be injected in the grid.

Based on $ \mathcal{U} $, the marginal contribution of an agent i can be expressed as $ \delta_{S}(i) = \mathcal{U}(S+\{i\}) - \mathcal{U}(S) $. A coalition S has thus an interest in adding an additional agent i if this marginal contribution is positive : 

\begin{equation}
\delta_{S}(i) \geq 0 \ \ \ \Leftrightarrow \ \ \ P_{S+\{i\}}^{CRCT \star} \geq P_{S}^{CRCT \star} \left( \dfrac{|S|+1}{|S|} \right)^{\alpha}
\label{eq:marginal_benefit}
\end{equation}

If $ \alpha $ is set to zero, agents are added as long as they increase the contract value of the coalition. If $ \alpha $ is greater than zero, additional agents have to increase the contract value by some factor. The utility function with $ \alpha $ is not necessarily convenient, here we relate $ \alpha $ to the mean sizes of the coalitions $ \bar{N} $. otherwise $ \mathcal{U}(S) $ tends to form coalitions of size approximately in the order of $ \bar{N} $, then :

\begin{equation}
\left[ \dfrac{\partial{ U}}{ \partial{|S|}} \right]_{|S| = \bar{N}} = 0
\label{eq:derivative}
\end{equation}

In order to get an estimator for $ \alpha $, we solve equation \ref{eq:derivative} in a Gaussian case (as in section \ref{sec:notations}). Furthermore, since considering all the possible interactions between agents is analytically intractable, we use here a mean approximation. Any quantity x that varies over the agent set is thus simplified in its mean value $ \bar{x} $. Solving equation \ref{eq:derivative} for $ \alpha $ in these conditions leads to:

\footnotesize
\begin{equation}
\alpha^{\star}_{\bar{N}} = \dfrac{0.7 \bar{\sigma}(\bar{\rho}-1)erf^{-1}(2 \phi - 1)}{\bar{\mu}\sqrt{\bar{N}(\bar{\rho}\bar{N}-\bar{\rho}+1)}+1.4 \bar{\sigma} erf^{-1} (2 \phi -1 ) (\bar{\rho}\bar{N}-\bar{\rho} + 1)} 
\label{eq:alpha_star}
\end{equation}
\normalsize

\begin{figure}
\includegraphics[scale=.6]{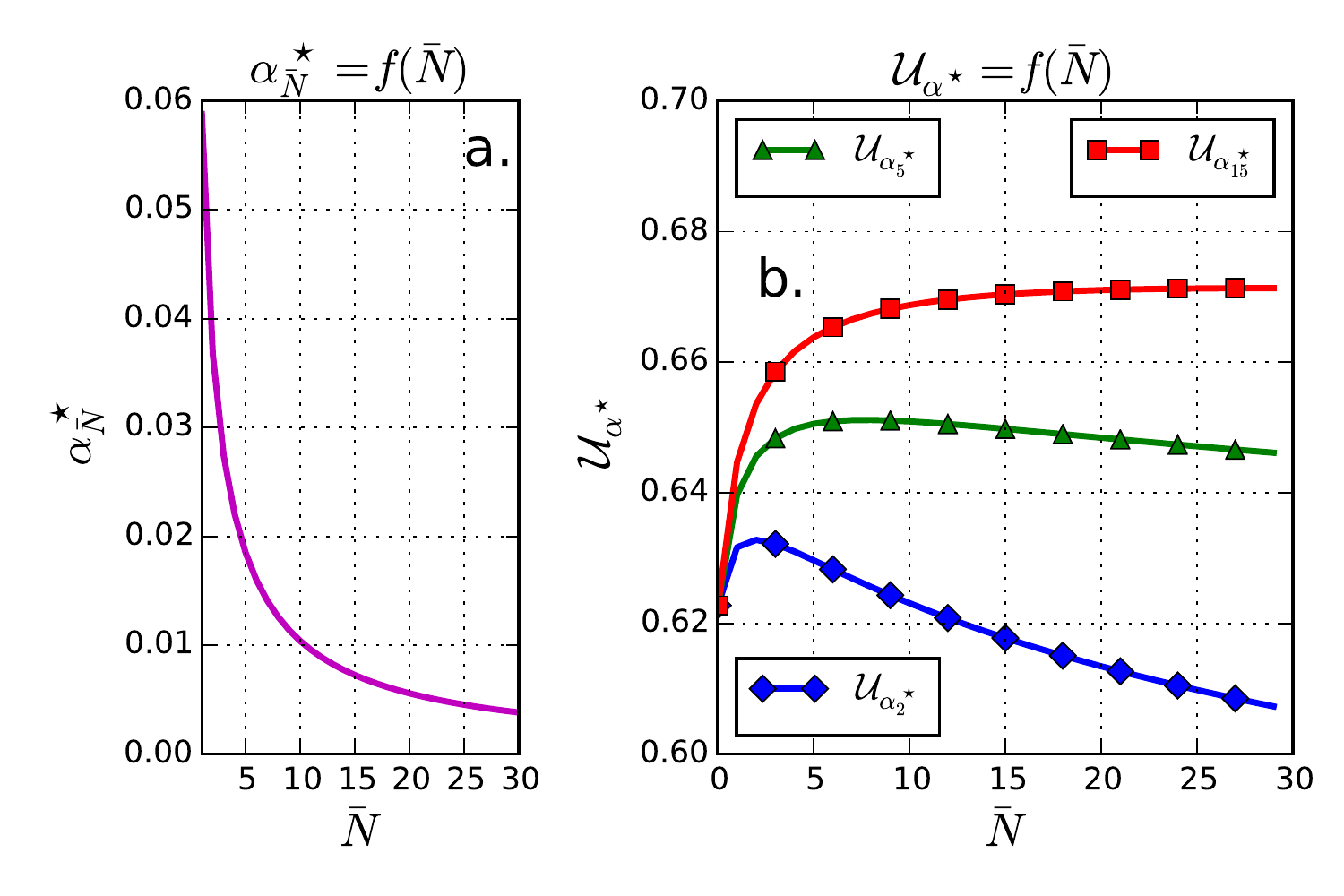}
\caption{Gaussian mean approximation. Subplot \textbf{a} shows how the parameter $ \alpha $ of the utility function should be chosen in function of the mean desired size of the coalitions (see equation \ref{eq:alpha_star}). Subplot \textbf{b} displays the corresponding utility functions for different values of $ \alpha $. Blue curve with diamonds favors very small coalitions of 2 agents while the green one with triangles favors 5 agents coalitions. Finally, the red curve with squares has an optimal size of 15 agents. }
\label{fig:mean_approx}
\end{figure}

Figure \ref{fig:mean_approx} shows how $ \alpha^{\star} $ and the utility function evolves according to the mean size of the coalitions $ \bar{N} $. These curves are only valid in the simplified Gaussian example considered here. Nonetheless, they will provide some guidance when using real data.

As can be pointed out, the purpose of $ \mathcal{U} $ is not to study coalitions stability against player defection which could be done through game theory, nor to redistribute the coalition's utility in terms of individual payoffs. But we aim to design $ \mathcal{U} $ as a measure of how good a given coalition is according to our criteria. In other terms does a given coalitions has a good production to risk ratio.

%
%
\section{Coalition Formation}
\label{sec:forming}
Section \ref{sec:notations} explained how contract values for the coalitions are computed, and in section \ref{sec:utility} we related this quantity to the utility and gains of a coalition. Since the computation time of this quantity is not negligible, we derive in this section the heuristic we used to form the coalitions structure. 

\subsection{Representing the correlation structure}
As seen in section \ref{sec:notations}, the variance of the aggregated production impacts directly the contract values, and depends on the covariances between the agents productions. We argue here that, by having some representation of the correlation structure between the agents, the search landscape for high utility coalitions could be reduced, such that good coalitions are more likely to be found quickly. Usually, this correlation structure is formalized with a covariance matrix or a correlation matrix that contains all the correlation coefficients between the agents : $ M = (\rho_{ij})_{\forall i,j \in \mathcal{A}^{2}}$. By using a metric to map this matrix in a weighted adjacency matrix (see section \ref{sec:related}), it is possible to obtain a graph representation of the correlation relationships between the agents. 

In the following, we use two opposite distance metrics for this mapping : 

\begin{equation}
\left\{ \begin{array}{lll}
			d_{ij}^{1} = 1 - \rho_{ij}^{2}, \\
			d_{ij}^{2} = \rho_{ij}^{2} = 1 - d_{ij}^{1}
\end{array} \right.
\end{equation}

Clearly, $ d^{1} $ (resp. $ d^{2} $) maps two correlated series as close points (resp. distant) while two uncorrelated series are distant (resp. close). These metrics enable us to compute a correlation graph $ G_{1} = (\mathcal{A}, E_{1}) $ and a "de-correlation" graph $ G_{2} = (\mathcal{A}, E_{2} ) $. For any i and j, the weight of the edge $ e_{ij} $ is $ d_{ij}^{1} $ in $ G_{1} $ and $ d_{ij}^{2} $ in $ G_{2} $.

In both cases, we want to keep only the edges which weights are located in the lower tail of the distance distributions. In other words, we want to compute the $ \epsilon$-graphs of $ G_{1} $ and $ G_{2} $ such that only meaningful edges remain. Selecting the right filter $ \epsilon $ is thus an important point since it affects the landscape search for the coalition formation. Unfortunately, there seems to be no clear consensus in the literature on how to select such a threshold. We will see later in this section that cliques in $ G_{2} $ are potential seeds for the coalitions. Since we want to generate $ N_{COAL} $ coalitions, we need at least $ N_{COAL} $ cliques of a given size to start. Besides, since we consider coalitions as disjoint, the starting cliques should be non overlapping. We thus select our optimal threshold for $ G_{2} $ as :

\begin{equation}
\label{epsilon_star}
\epsilon^{\star} = min_{ \epsilon \in [0,1]} \left\{ \epsilon\ s.t.\ |\Theta_{k}(G_{2}^{\epsilon})| \geq N_{COAL} \right\}
\end{equation} 
where $ G_{2}^{\epsilon} $ is the de-correlation graph $ G_{2} $ filtered by $ \epsilon $, and $ \Theta_{k}(G) $ is the set of non overlapping cliques of size k in a given graph G. In other words we select $ \epsilon^{\star} $ as the smallest threshold possible such that the filtered de-correlation graph contains at least $ N_{COAL} $ non overlapping cliques of size k. The existence of $ \epsilon^{\star} $  as defined in equation  \ref{epsilon_star} is not guaranteed. The users has indeed to provide consistent values of $ N_{COAL} $ or k compared to the size of the agent population $ \mathcal{A} $. 

\subsection{Cliques}

In \cite{Garas2008} the structural roles of weak and strong links on financial correlation graphs is investigated. The author shows that strong links, accounting for strong correlation relationships, are responsible for the clustering, while weak links provide the connectivity between clusters. Indeed, if we consider three items, say a, b, and c such that a and b are strongly correlated and b and c are also strongly correlated, then it is likely that a and c are also strongly correlated. It can be easily shown using the cosine addition formula\footnote{ $ cos(a+b) = cos(a)cos(b) - sin(a)sin(b) $ }, that if $\rho_{ab} > x $ and $\rho_{bc} > x $ with $ x>0 $, then $\rho_{ac} > 2x^{2}-1$). Correlation graphs capture this weak transitivity notion through clusters of correlated series.

Nevertheless de-correlation seems like a more complex concept than correlation in the sense that there is not even a partial notion of transitivity when it comes to it. Therefore, the clustering coefficients of $ G_{1}^{\epsilon} $ is much higher than the one of $ G_{2}^{\epsilon} $. This can be seen as another formulation of \cite{Garas2008} on the structural roles of weak and strong links on financial correlation graphs. Strong links, accounting for strong correlation relationships, are responsible for the clustering, while weak links provide the connectivity between clusters. Searching for clusters in $ G_{2}^{\epsilon} $ and hoping that this strategy will provide a nice coalition structure of internally uncorrelated coalitions seems thus pointless.

Consider now a clique in  $ G_{2}^{\epsilon} $, which is a complete subgraph of $ G_{2}^{\epsilon} $. This is indeed a structure of interest for our purpose. Since there is a link for every pairs of nodes, we know, by construction, that a clique has a mean correlation and a maximum correlation less than $ \epsilon $. 

\begin{figure}
\includegraphics[scale=.6]{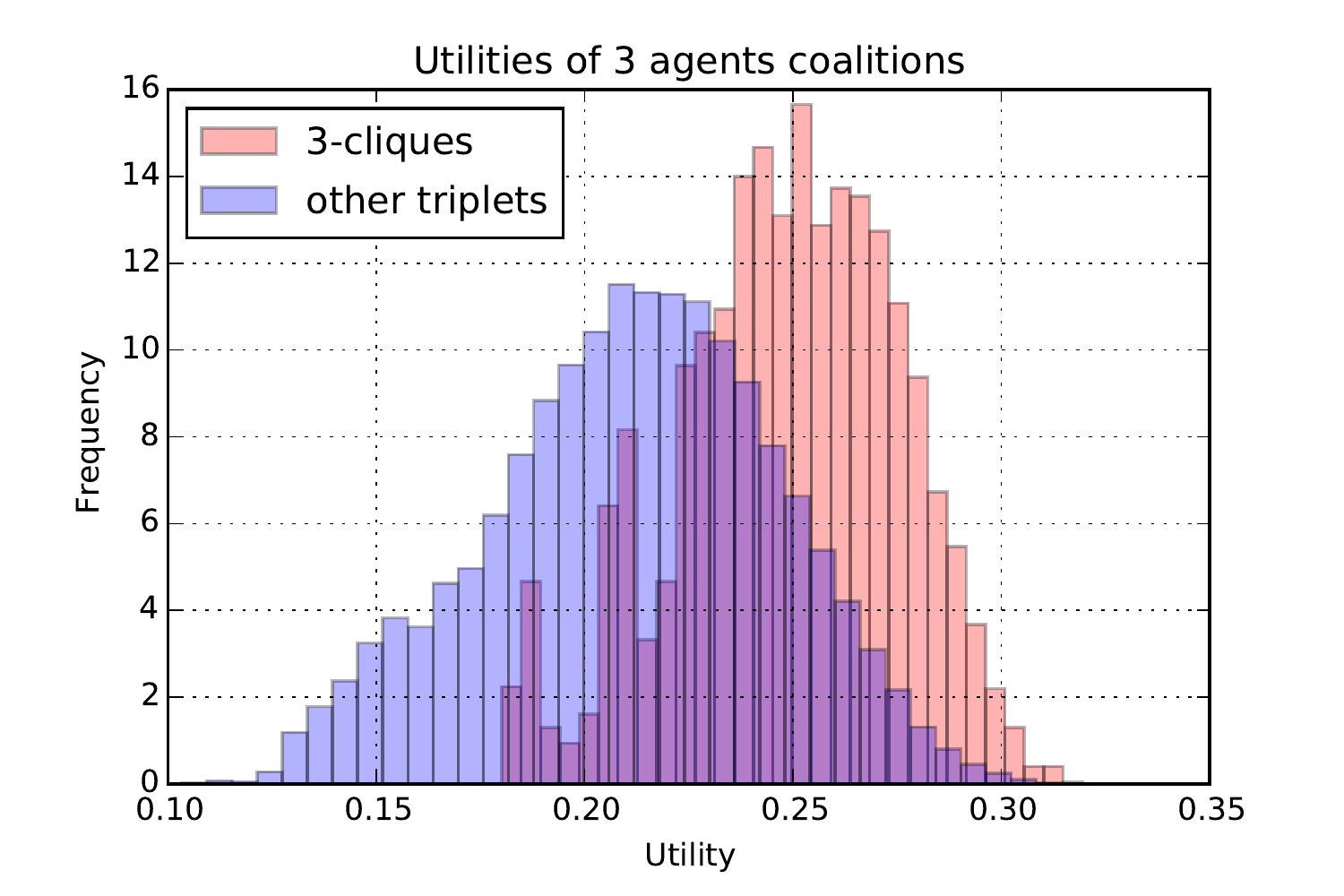}
\caption{Histograms of utility values for coalitions of size 3. Red bars stands for cliques in the decorrelation graph, and blue bars represents all the other possible triplets. As visible, cliques tend to exhibit higher utilities than randomly selected coalitions.}
\label{fig:histo_cliques}
\end{figure}

Figure \ref{fig:histo_cliques} shows the distributions of the utility values for cliques of size 3 (triangles) in $ G_{2}^{\epsilon^{\star}} $ and for all the other possible triplets of agents. It is clearly visible that cliques tend to exhibit higher utilities because of their de-correlation property. Choosing cliques in $ G_{2}^{\epsilon^{\star}} $ as coalitions seems therefore appealing. Nevertheless, the quality of the results seems to decrease as the sizes of the cliques increase. Indeed, the larger the desired cliques, the more dense $ G_{2}^{\epsilon^{\star}} $ becomes (see equation \ref{epsilon_star}). There is a point where cliques results more from noisy edges than true de-correlation, which decreases the quality of the results.

Directly mapping cliques to coalitions by this de-correlation oriented approach is thus not sufficient. It is indeed possible that adding agents to these cliques has the combined effect of increasing the expected production while decreasing its stability. The question revolves around measuring the benefits of this production surplus compared to the disadvantage of having coalition with high volatility. This can be quantified by the marginal benefit in equation \ref{eq:marginal_benefit}.

\subsection{Algorithm}
The algorithm takes inputs from :

\begin{itemize}
	\item \textbf{The agents :} historical series of available productions $P_{i}$, 
	\item \textbf{The grid operator :} market entrance policy $ (P^{MIN},\phi) $,
	\item \textbf{The "user" :} Number of desired coalitions $ N_{COAL} $ and size of starting cliques k.
\end{itemize} 

The first steps consists in computing the de-correlation graph $ G_{2} $ as well as the optimal threshold $ \epsilon^{\star} $. Cliques of size k in $ G_{2}^{\epsilon^{\star}} $ are considered as coalition seeds. The next step is a local greedy improvement over the landscape represented by  $ G_{2}^{\epsilon^{\star}} $. Cliques add alternatively the node $ i^{\star} $ in their neighborhood that yields the best marginal benefit $ MAX_{ i \in N(clique) } \delta_{clique}(i) $ where $ N(clique) $ is the neighborhood of a given clique. This addition occurs only if $ i^{\star} $ is not already involved in another coalition, and if $ \delta_{clique}(i^{\star}) \geq 0 $, meaning that utilities are increasing. The algorithm stops when all nodes are distributed in a coalition or when the global utility stops increasing. See the details in algorithm \ref{alg:algo1} in the appendix.

%
%
\section{Results}
\label{sec:results}

The algorithm presented in the previous section is supposed to generate a given number of coalitions that have good utilities. As it comprises mainly of a greedy optimization based on local improvements, there is no guarantee that the algorithm finds the global optimum. Since there is, to our knowledge, no state of the art algorithm that aggregates uncorrelated agents in an optimum way (see section \ref{sec:related} for related problems), we compare the results with :

\begin{itemize}
\item \textbf{Random sampling of coalitions :} Coalitions are formed randomly without any other constraint that the desired size. This enables us to have an idea about the distributions of utility values for coalitions of a given size.
\item \textbf{Random sampling of coalition structures :} Coalition structures are sampled randomly by shuffling and random divisions of the agents. Algorithm \ref{alg:algo2} uses such a sampling and returns the highest utility coalition structure sampled. This algorithm will be refered to as "\textit{random}" in the following.
\item \textbf{Correlated :} This is the complete opposite of our algorithm. It uses the correlation graph $ G_{1} $ and performs a community detection. The resulting coalitions have thus very high internal correlations. We thus expect this algorithm to perform very bad compared to the others. See algorithm \ref{alg:algo3}.
\end{itemize}

\begin{figure}
\includegraphics[scale=.6]{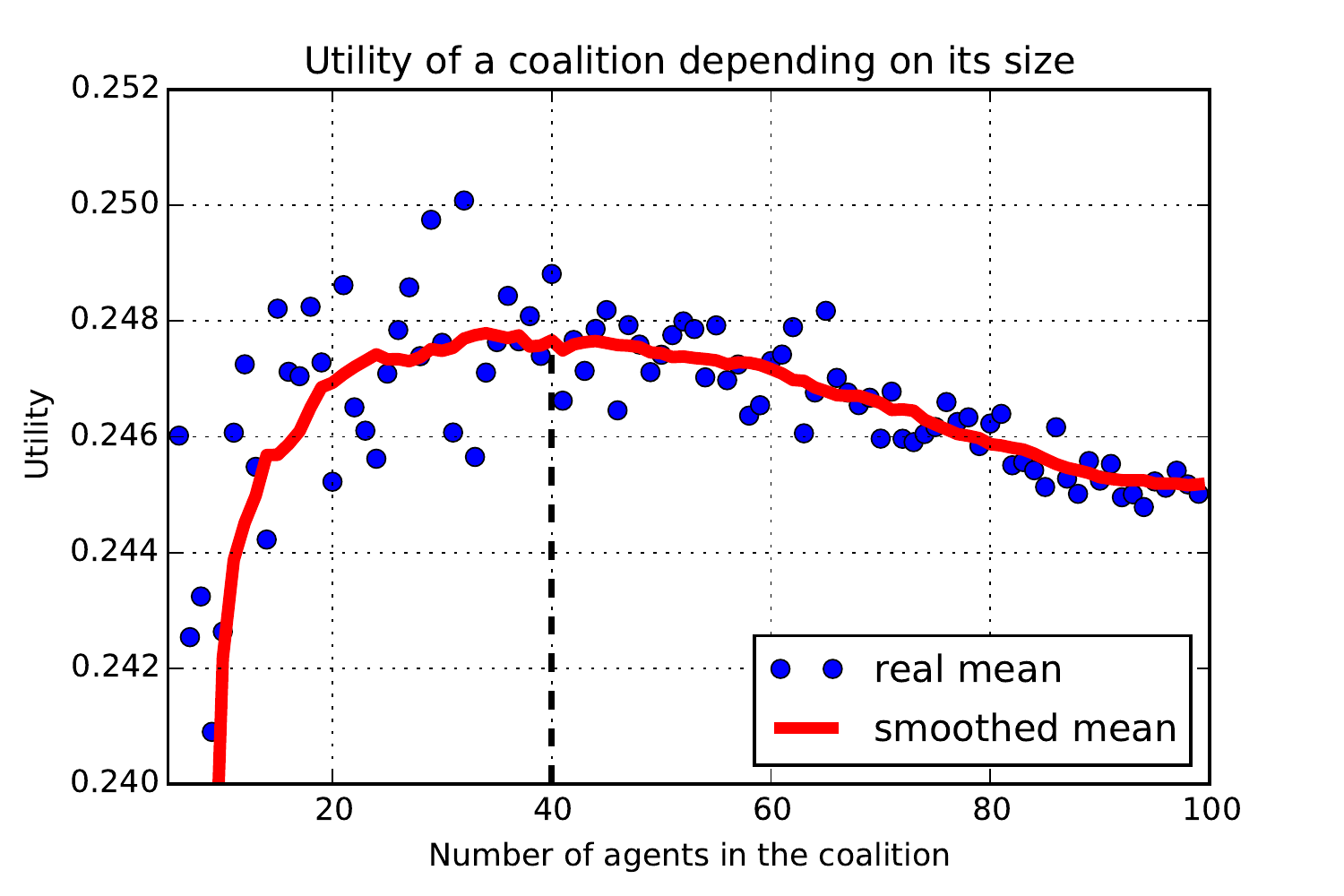}
\caption{Utility of random coalitions depending on their size. Blue dots show real mean utility values and the thick red curve its smoothed version by applying a Savitzky-Golay filter. On this plot the $ \alpha $ parameter of the utility function was selected according to equation \ref{eq:alpha_star} in order to favor 40 agents coalitions.}
\label{fig:real_utility2}
\end{figure}

Before running the algorithms, we need to calibrate the utility function by choosing the value of the $ \alpha $ parameter. Recall that the purpose of this parameter is to take into account some constraints on the coalition's sizes if needed. In this paper, neither the communication network nor the electrical grid are explicitly considered. Thus, we do not have any technical constraints on coalition sizes even if we designed the utility such that these could be taken into account. We select the desired size as being $ \lfloor N/N_{COAL} \rfloor $ (where $ \lfloor.\rfloor $ means floor). Figure \ref{fig:real_utility2} shows how the mean utility of a coalition evolves with its size when the optimum size is set to 40 agents. Using equation \ref{eq:alpha_star} to estimate $ \alpha $ based on the mean quantities and Gaussian approximations seems to give acceptable results for the utility function behavior.

\begin{figure}
\includegraphics[scale=.6]{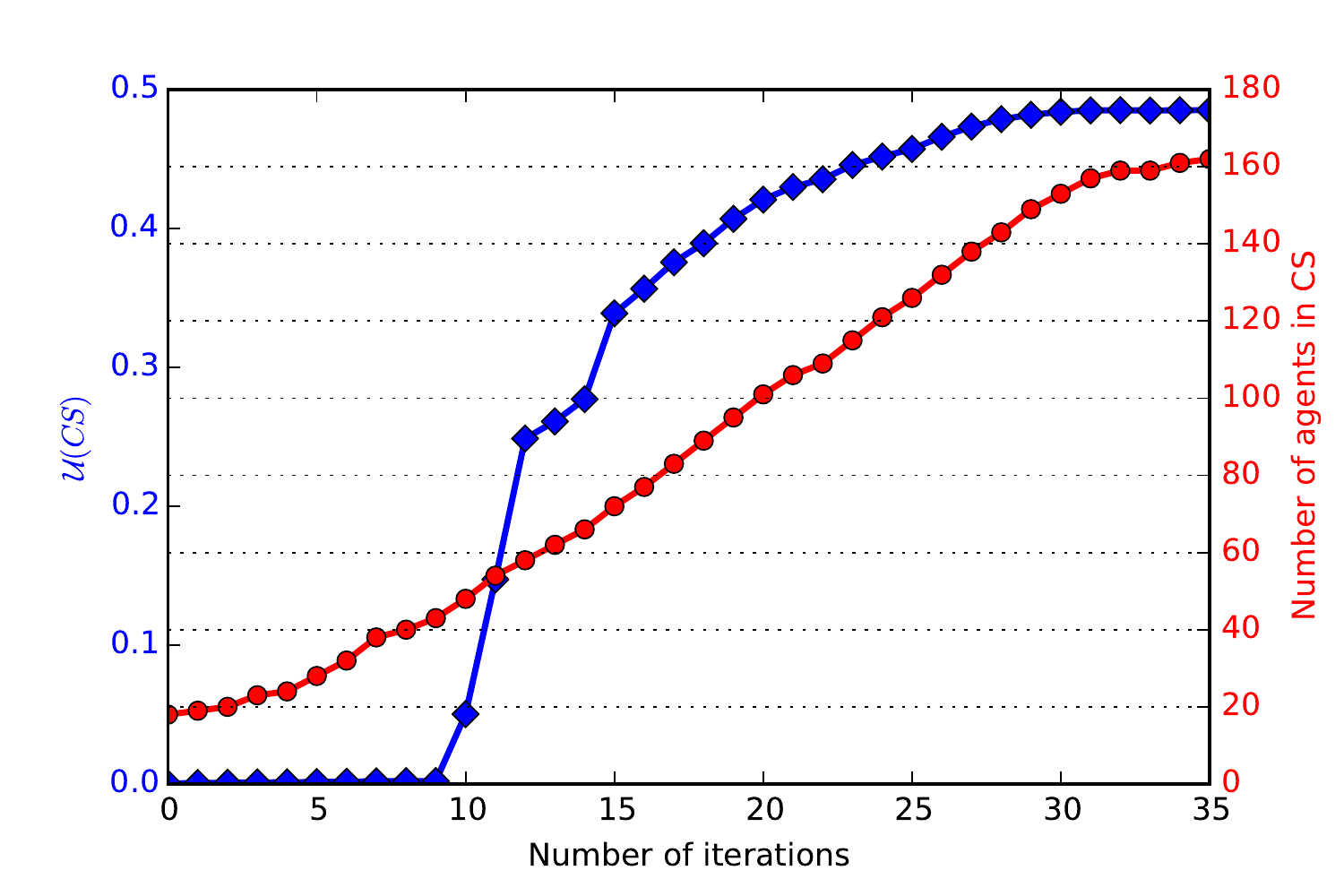}
\caption{Evolution of the global utility $ \mathcal{U}(CS) $ (blue diamond curve, left axis) and the number of agents involved in the coalitions (red circle curve, right axis) during the greedy optimization of algorithm \ref{alg:algo1} }
\label{fig:search}
\end{figure}

Figure \ref{fig:search} displays the evolution of the global utility and the number of involved agents during the course of the greedy algorithm \ref{alg:algo1}. The transition from an invalid to a valid coalitions is clearly visible on the blue diamond curve and occurs between iteration 10 and 15. After this transition, coalition's utilities improve slowly up to a maximum point. 

\begin{figure}
\includegraphics[scale=.6]{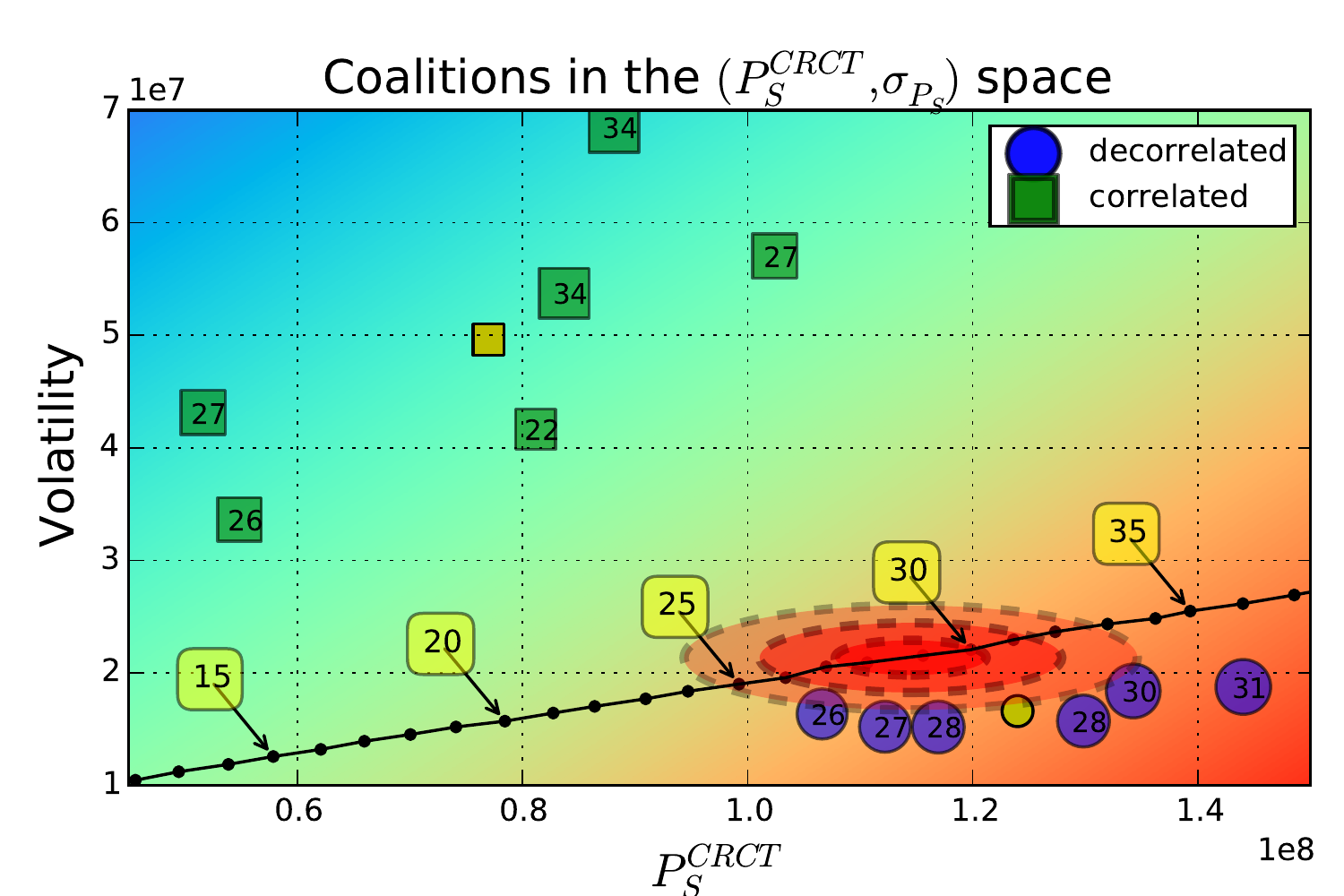}
\caption{Coalitions formed in the (contract value, volatility) space. The color map indicates qualities of portions of the plane. The closer to red the better (high contract values with small volatility). On the opposite, blue areas show poor quality (small contract values with high volatility). Blue dots stand for the decorrelated coalitions that we formed while green squares show correlated coalitions. The smaller yellow markers stand for the gravity centers of the coalition structures. The black dotted line shows how contract values and volatility evolve  when the size of the coalitions increases (a subset of the points are labeled by the size of the coalition they represent). Each point is the average over $ 10^{5} $ unconstrained draws of a random coalition. As we can also draw random coalition structures, we show the distribution of their gravity centers by the red ellipses (center is the mean, and each ellipse corresponds to one  standard deviation). For instance in our algorithm the utility function favors balanced coalitions. We also loosely constrained the sizes of the coalitions in the utility function. }
\label{fig:coalitions}
\end{figure}

Figure \ref{fig:coalitions} shows the coalitions formed with the considered algorithms in the contract value / volatility space. The color map in the background indicates regions where we expect high utilities (red) and the ones where we expect very poor utility values (blue). The bottom right corner, with high contract values and low volatilities, is therefore the region where we wish to form our coalitions. A single coalition is represented by a marker and the color and shape of a marker indicates by which algorithm the coalition has been formed. Besides, the sizes of the coalitions are indicated on the markers, and the marker size is also proportional to the coalition size. We can see that the utility function results in approximatively balanced coalitions. Small yellow markers indicates the gravity centers of their respective coalition structures. The coalitions of correlated agents (green squares) are clearly of poor quality according to our criteria since they can only afford small production contracts, and with a very high volatility. 

On figure \ref{fig:coalitions}, the decorrelated coalitions (blue dots) are closer to the bottom right corner indicating a much better quality in term of productivity over volatility ratio. The black dotted line indicates the mean values for the random coalitions sampling technique. Each small dot stands for the mean position of all sampled coalitions of this given size. Variances are not indicated for readability, but are usually quite large since this sampling only takes the size as a constraint. We can see that as coalitions get larger, they tend to increase on average their contract values, but at the price of a higher volatility. 

On figure \ref{fig:coalitions}, the results of the random coalition structure sampling are shown with the red ellipses that represent the distribution of the gravity centers of the sampled structures. Since the center of the ellipses stands for the mean and each ellipse adds one standard deviation, more than $ 99\% $ of the sampled gravity centers are within the largest ellipse. The small yellow dot below the ellipses indicates the gravity center of our solution. It is thus visible that our greedy graph based algorithm is able to find a quite good coalition structure in terms of volatility and contract values.

A key point for the coalitions, besides stability and productivity, is their resilience. The resilience of a system can be roughly described as its ability to perform its tasks when subject to failures of its components. Therefore, the notion of resilience we will use in the following can be seen as the ability of the coalition structures to inject stable power in the grid when node failures occur. According to our model, the grid operator specified two thresholds ($P^{MIN}$ and $ \phi $) such that the power injected by every coalition is constrained : $ P_{S}^{CRCT} \in [P^{MIN}, P_{S}^{CRCT \star}] $. As long as a coalition can propose a contract value higher than $ P^{MIN} $, it is valid and allowed to enter the energy market. We define the resilience of a coalition S as the probability that S produces more than the $ P^{MIN} $ threshold :

\begin{equation}
\mathcal{R}_{S} = Pr[P_{S} >= P^{MIN}] = 1 - Pr[P_{S} < P^{MIN}]
\end{equation}

And we extend this measure to the coalition structures :

\begin{equation}
\mathcal{R}_{CS} = \prod_{S \in CS} \left( 1 - Pr[ P_{S} < P^{MIN} ] \right)
\label{eq:resilience}
\end{equation}

\begin{figure}
\includegraphics[scale=.6]{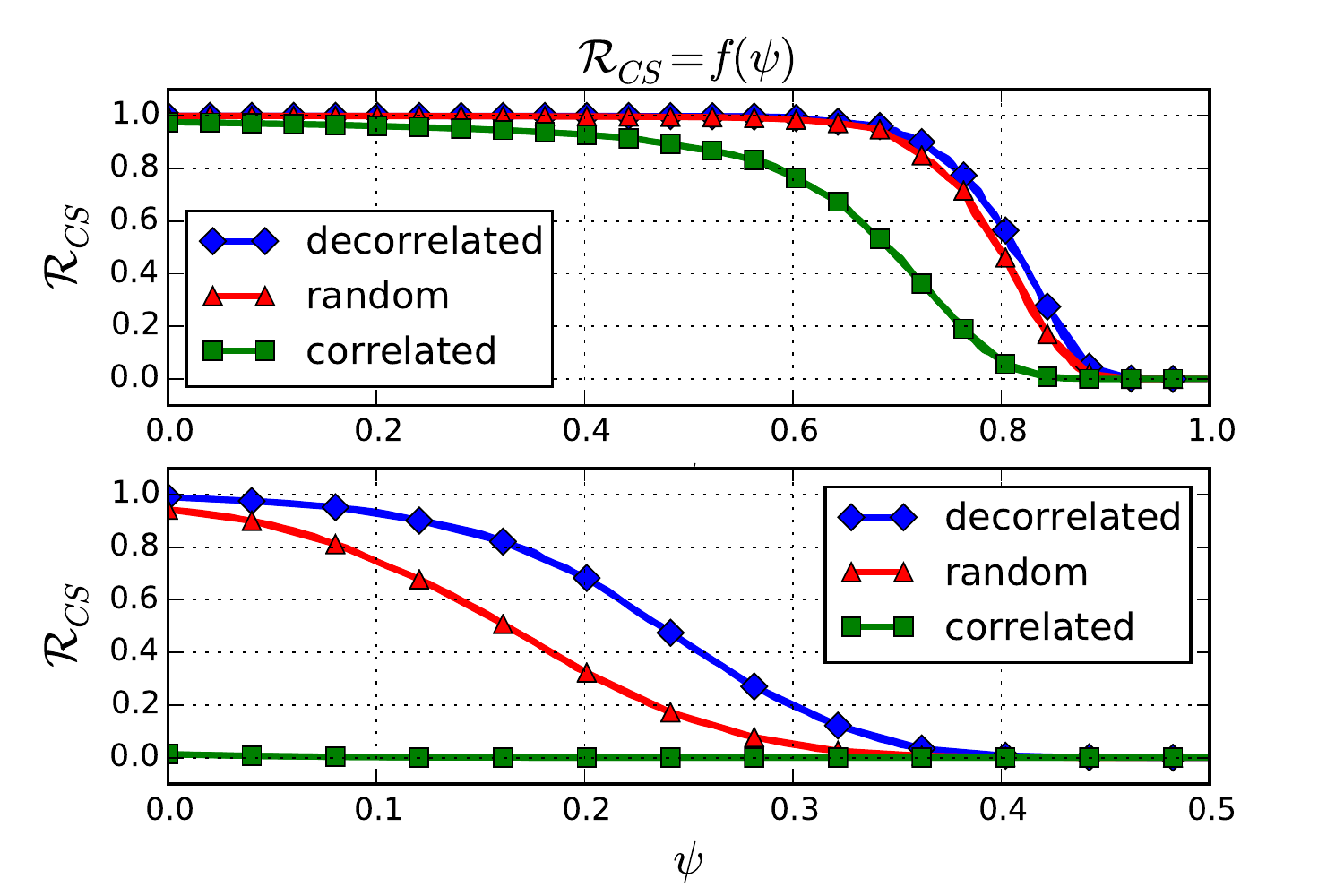}
\caption{Resilience of the coalition structures when nodes fail randomly (see equation \ref{eq:resilience}) for $ P^{MIN} = 10MW $ (top subplot) and $ P^{MIN} = 80MW $ (bottom subplot) }
\label{fig:resilience}
\end{figure}

We consider that prosumers fail randomly, and we denote by $ \psi \in [0,1] $ the fraction of agents that failed. Figure \ref{fig:resilience} exhibits how the resilience of the coalition structures evolves according to $ \psi $. On the top subplot, $ P^{MIN} $ was voluntarily selected relatively low such that the resiliences of the three structures fit on the same figure. When the $ P^{MIN} $ requirement increases, the differences between the algorithms also increase as visible on the bottom subplot of figure \ref{fig:resilience}. The decorrelated coalitions seem to achieve a more resilient production on the market in the sense that they are able to sustain a higher fraction of node failures.

%
%
\section{Conclusion}
\label{sec:conclusion}

In this paper we studied how aggregations of prosumers could be authorized to sell their surplus of production to the grid operator. By relying on the past values of the agents, we constrained the market entry to both sufficiently productive and stable coalitions. The power that a coalition is able to propose on the market is therefore related to production and stability. As the correlations between the prosumers that form these coalitions impact directly their volatilities, we seek uncorrelated aggregations of agents. We used a graph representation of the correlation relationships between the agents as a reduced landscape for the coalition formation. A greedy algorithm that starts with cliques of the "de-correlation" graph of the agents and makes local improvements offers a good compromise between speed and quality of the results. We compare these results with random samplings, and an opposite strategy that clusters correlated agents together. We showed that the coalitions resulting from our algorithm are able to provide more power to the grid with a lower volatility. Because they tend to have globally a better production over volatility ratios, these coalitions will tend to use less storage and waste less energy than more unstable coalitions. We plan to study these benefits for the control of the aggregations in future works.

Because in real situations, agents are prone to failure, resilience is also an important criterion for the quality of the aggregations. We therefore studied how the coalitions are able to remain on the market when their agents fail randomly one by one. We showed that, in this situation, the coalitions resulting from our algorithm better withstand losses of agents.

%
%

\bibliographystyle{IEEEtran}  
\bibliography{Article}

\appendix
\subsection{Algorithms}

\begin{algorithm}
 \KwData{$P_{i}$ series,\\ Grid policy $ (P^{MIN},\phi) $,\\ Desired number of coalitions $ N_{COAL} $,\\ size of starting cliques k}
 \KwResult{ $ CS = \{ S_{1},...,S_{N_{COAL}}\} $ }
 Compute $ G_{2}^{\epsilon^{\star}} $ \;
 Find the $ N_{COAL} $ cliques in $ G_{2}^{\epsilon^{\star}} $\;
 \While{$ \mathcal{U}(CS) $ is improving}{
 	\For{each clique}{
 		Find $ i^{\star} $ \;
 		\If{ $ \delta_{clique}(i^{\star}) \geq 0 $ }{
 			$ clique \leftarrow clique \cup \{i^{\star} \} $ \;
 			}
 		\If{ $ \exists j \in clique,\ s.t\ \delta_{clique}(j) < 0 $}{
 			$ clique \leftarrow clique - \{j \} $ \;
 			}
   		}
  	}
 \caption{Local greedy optimization algorithm}
 \label{alg:algo1}
\end{algorithm}

\begin{algorithm}
 \KwData{Agent set $\mathcal{A}$,\\ Desired number of coalitions $ N_{COAL} $, \\ Maximum number of iterations $ Loop^{max} $}
 \KwResult{ $ CS = \{ S_{1},...,S_{N_{COAL}}\} $ }
 $ N^{loop} \leftarrow 0 $ \;
 $ CS^{\star} \leftarrow \emptyset $\;
 \While{$ N^{loop} < Loop^{max} $}{
 	$ CS \leftarrow SelectRandomCS( ) $\;
 	\If{ $ \mathcal{U}(CS) > \mathcal{U}(CS^{\star}) $ }{
 		$ CS^{\star } \leftarrow CS $\;
 		}
 	$ N^{loop} \leftarrow  N^{loop} + 1 $\;
  	}
  return $ CS^{\star} $
 \caption{Random algorithm}
 \label{alg:algo2}
\end{algorithm}

\begin{algorithm}
 \KwData{$P_{i}$ series,\\ Desired number of coalitions $ N_{COAL} $,\\ search step size $ \beta << 1 $}
 \KwResult{ $ CS = \{ S_{1},...,S_{N_{COAL}}\} $ }
 $ \epsilon \leftarrow 1 $ \;
 $ CS \leftarrow \emptyset $\;
 \While{$ |CS| < N_{COAL} $}{
 	Compute $ G_{1}^{\epsilon} $ \;
 	$ CS \leftarrow computeClusters( G_{1}^{\epsilon} ) $\;
 	\If{ $ |CS| = N_{COAL} $ }{
 		return CS\;
 		}
 	\Else{ 
 	$ \epsilon \leftarrow \epsilon - \beta $\;
 	}
  	}
 \caption{Correlated algorithm}
\label{alg:algo3}
\end{algorithm}

\subsection{Net production series}\label{appendixB}

Data were collected from \cite{Infoclimat} (similar data can be found at \cite{NCDC} for the United States). The variables used in the simulation are :

\begin{itemize}
\item Average wind speed (in $m.s^{-1}$)
\item Nebulosity (integer in $[0,8] $)
\item Temperature (in degree Celsius)
\end{itemize} 

\subsubsection{Wind power curve}
Power curves are functions that, for a given type of generator, map some input quantity to the output power produced. For wind-turbines and solar arrays these functions are well studied and approximations have been proposed \cite{Lydia2014} \cite{Piedallu2007, Piedallu2008}. For the wind turbines, the power curve can be specified by 4 values :

\begin{itemize}
\item Cut-in-speed : The wind speed at which the turbine first starts to rotate and generates power.
\item Rated-output-power : The maximum power that the turbine can generate.
\item Rated-output-speed : The wind speed at which the turbine attains its rated output power.
\item Cut-out-speed : The speed at which the turbine is turned off as not to damage the rotor.
\end{itemize}

The most interesting part is the increase of output power when the wind speed is in the cut-in-speed rated-output-speed range. Even if sometimes a simple linear model is used, the increase has been shown to be non linear and some more complex exponential fit can be found in the literature \cite{Lydia2014}.

\subsubsection{Solar power curve}
The input quantity desired for our power curve model for solar arrays is a radiance in $ W.m^{-2} $, which can be difficult to find in weather station available data. As we mainly collected nebulosity series, we used the Helios model described in \cite{Piedallu2007, Piedallu2008}. This model enabled us to compute perfect (clear blue sky situation) solar radiances at some specific locations on earth and at given timestamps. As nebulosity is a measure of the sky cloudiness, we can use the nebulosity series as degradation factors on the clear blue sky model (see \cite{Piedallu2007, Piedallu2008} for more details) :

\begin{equation}
\left\{ \begin{array}{lll}
			\Psi{real}(t) = \Psi_{perfect}(t) \eta(t) \\
			\eta(t) = 1-0.75 \left( \dfrac{N(t)}{8} \right)^{3.4}
\end{array} \right.
\end{equation}
where $ \Psi_{perfect}(t) $ and $ \Psi{real}(t) $ are respectively the clear blue sky and real radiances at time t, $ \eta(t) $ is the degradation factor at time t, and $ N(t) $ is the nebulosity index at time t.

Once we have input data in the forms of radiances, we compute the production of a solar array with the following simplified power curve :

\begin{equation}
\mathcal{F}_{PV}(\Psi{real}(t)) =  S_{PV} \Psi{real}(t)  e_{PV}
\end{equation}
where $ S_{PV} $ is the surface of the array, and $ e_{PV} $ is its efficiency. The very simple form of this power curve is due to some simplifications in order not to overload the model. For instance, it does not take into account angles and orientations degradations. These could be incorporated if needed by changing the power curve in the simulations.

\subsubsection{Consumption}
Modeling electric consumption has already been widely tackled in the literature. Models can be basically divided into two main categories : Top-down and bottom-up approaches. Top-down techniques take aggregated consumption data as inputs and try to estimate individual consumption patterns while bottom-up methods use a fine modeling of users consumptions as to obtain realistic aggregated consumption curves. In this paper, we used a bottom-up model since the end user, or relatively small aggregations of end users, are in our interest. The main objective was to capture both daily patterns and seasonal variations of the consumptions. We assumed an additive model where the consumption of an agent is the sum of a seasonal heating term that depends on the outside temperature and an electronic consumption term that only depends on the hour of the day. By denoting $ \tau(t) $ the outside temperature at timestamp t, we can express the consumption $ P_{i}^{D}(t) $ of agent i at time t :

\begin{equation}
P_{i}^{D}(t) = \mathcal{F}_{i}^{heat}(\tau(t), t) + \mathcal{F}_{i}^{elec}(t)
\end{equation}
where $ \mathcal{F}_{i}^{heat}(\tau(t), t) $ is the power curve that maps the temperature to a heating consumption, and $ \mathcal{F}_{i}^{elec}(t) $ computes the consumption of agent i (other than heating) at a given hour of the day. In the simulation, all agents have a desired inside temperature $ T_{i} $, supposed to be a constant for simplification. By using thermodynamic laws $ \mathcal{F}_{i}^{heat}(\tau(t), t) $ can be approximated by :

\begin{equation}
\mathcal{F}_{i}^{heat}(\tau(t), t) = \dfrac{B_{i}}{R_{i}} \left[ T_{i} - \tau(t) \right]
\end{equation}
where $ B_{i} $ is the surface of thermal exchanges for agent i and $ R_{i} $ is their thermal resistance.

We denote by $ \Omega_{i} $ the maximum consumption possible for agent i, which is basically the sum of all its appliances powers. We also denote by $ \omega_{i}(t) = \{ \omega_{i}(t_{0}),...,\omega_{i}(t_{24}) \} $ the vector of the average fraction of $ \Omega_{i} $ used for each hour. We can therefore write :

\begin{equation}
\mathcal{F}_{i}^{elec}(t) = \Omega_{i} ( \omega_{i}(t) + \epsilon )
\end{equation}
where $ \epsilon $ is a noise term. The vector $ \omega_{i}(t) $ enables us to easily differentiate agent consumption behaviors. Business or residential areas for instance can be easily distinguished with this kind of model.

\end{document}